\documentclass[twocolumn,amsmath,amssymb,nobibnotes, aps,pra]{revtex4-1}
\usepackage[T1]{fontenc}
\usepackage[latin9]{inputenc}
\usepackage{geometry}
\usepackage{graphicx}
\usepackage{calc}
\usepackage{amsfonts}
\usepackage{amsmath}
\usepackage{extarrows}
\usepackage{bm}
\usepackage{graphicx}
\usepackage{overpic}
\usepackage{amsmath,mathrsfs,dcolumn,dsfont}
\usepackage{amssymb}
\usepackage{epsfig,epstopdf}
\usepackage{bm,bbm,mathrsfs}
\usepackage{threeparttable}
\usepackage{comment}
\usepackage{ulem}
\usepackage[large]{subfigure}
\usepackage{color}

\begin{document}
	\title{Monitoring Spontaneous Charge-density Fluctuations by Single-molecule Diffraction of Quantum Light}

\author{Konstantin E. Dorfman}
\email{Email: dorfmank@lps.ecnu.edu.cn}

\affiliation{State Key Laboratory of Precision Spectroscopy, East China Normal
University, Shanghai 200062, China}

\author{Shahaf Asban}
\email{sasban@uci.edu}

\affiliation{Department of Chemistry and Department of Physics and Astronomy,
University of California, Irvine, California 92697-2025, USA}

\author{Lyuzhou Ye}

\affiliation{Department of Chemistry and Department of Physics and Astronomy,
University of California, Irvine, California 92697-2025, USA}

\author{Jérémy R. Rouxel}

\affiliation{Laboratory of Ultrafast Spectroscopy, École Polytechnique Fédérale
de Lausanne, CH-1015 Lausanne, Switzerland}

\affiliation{SwissFEL, Paul Scherrer Institut, 5232 Villigen PSI, Switzerland}

\author{Daeheum Cho}

\affiliation{Department of Chemistry and Department of Physics and Astronomy,
University of California, Irvine, California 92697-2025, USA}

\author{Shaul Mukamel}
\email{smukamel@uci.edu}

\affiliation{Department of Chemistry and Department of Physics and Astronomy,
University of California, Irvine, California 92697-2025, USA}

\date{\today}%

\begin{abstract}
Homodyne X-ray diffraction signals produced by classical light and
classical detectors are given by the modulus square of the charge density
in momentum space $\left|\sigma(\mathbf{q})\right|^{2}$, missing
its phase which is required in order to invert the signal to real
space. We show that quantum detection of the radiation field yields
a linear diffraction pattern that reveals $\sigma(\mathbf{q})$ itself,
including the phase. We further show that
repeated diffraction measurements with variable delays constitute
a novel multidimensional measure of spontaneous charge-density fluctuations. Classical diffraction, in contrast, only reveals
a subclass of even-order correlation functions. Simulations of two dimensional signals obtained by two diffraction  events are presented
for the amino acid cysteine.
\end{abstract}




\date{\today}
\maketitle


\section{Introduction}

Photon counting, as described by the quantum theory of detection,
is associated with annihilation of a radiation mode \cite{gla63}.
Any detectable change in the number of photons requires at least two light-matter interactions. Diffraction of a classical source on quantum matter is thus a second order
process in the light/matter interaction. Sources with a low photon
flux \cite{hon87,bra00,tro16,kal16,lee16,nor18,pat18} or short wavelength \cite{kra09,ish12,cor13,chi14,bos16} - that can detect $\left(\Delta n\right)\hbar\omega$
is detectable ($\Delta n$ being the change in photon number) - now
exist.Taking the quantum nature of light into account is now called for.

Multidimensional diffraction can be measured by
photon coincidence counting obtained by subjecting the
molecule to sequences of pulses. The underlying matter information is given by the multi-point correlation functions of the
charge density which governs the spontaneous charge
fluctuations. The response and spontaneous fluctuations of
both field and charge density are  mixed due to their quantum nature and classical response theory, which is causal does not apply \cite{coh03}. Thus, multidimensional spectroscopy, which involves several
perturbations followed by a single measurement is fundamentally different from multidimensional diffraction, which consists of a series of
measurements, and
thus may not be retrieved simply by data processing of classical
signals. Multidimensional diffraction carries new type of
information related to spontaneous charge fluctuations, which is not accessible by classical light \cite{dor15}.

 In this letter we consider off-resonant diffraction of
nonclassical X-ray sources, and explore phase dependent quantum corrections to diffraction, involving a \textit{single} light-matter interaction.
Photons are not generated in this order (this requires two  interactions), which only causes phase change of the field. This results in a detectable photon intensity
diffraction pattern when coupled to local quantum fluctuations at
the detector. We denote this process as linear quantum diffraction (LQD) (i.e linear in the charge density). 

We consider an incoming
light prepared either in a coherent state or in a Fock state interacting
with a local field mode which is eventually detected by photon annihilation in the detected mode \cite{gla63}. Field intensity
measurements show that local quantum fluctuations at
the detector coupled to the detected mode generate signal linear in the charge density. Coherent (classical-like)
or single-photon states provide higher degrees of spatial and spectral
resolution, whereas an $N$- photon Fock state yields lower resolution.

Crystallographic signals generated by classical
light are quadratic in the charge density in momentum space
$\sigma(\mathbf{q})$. The phase is not available and phase reconstruction
algorithms \cite{miao1999extending, PhysRevB.67.174104} or heterodyne detection \cite{PhysRevLett.120.243902} are required to retrieve the real-space
charge density $\sigma(\mathbf{r})=\int d\mathbf{q}e^{i\mathbf{q}\cdot\mathbf{r}}\sigma(\mathbf{q})$. Heterodyne detection
of the signal field is achieved by interference with a Local Oscillator
(LO) \cite{mar08}, which must be varied for each scattering angle. Phase reconstruction algorithms usually require a reasonable initial
guess in order to converge to the correct structure \cite{maiden2009improved, candes2015phase}. Signals linear
in the charge density, can reveal the phase of the Fourier-transformed
charge densities and the crystallographic image. Thus quantum
detected diffraction offers an interesting possibility for overcoming the
phase problem without scanning the LO for each detection angle. Furthermore,
classical diffraction can be viewed as an ensemble average of  different trajectories.
Each detection event results from a trajectory terminated in a point at the
detector. It is further blurred by the detector response function,
even for infinitesimal detection area (pixel size). Using quantum detection,
this response can be studied at the single trajectory level,
 enhancing the resolution by reducing the spread and minimizing the noise \cite{asb18}.

Repeated measurements involving sequences of n delayed pulses result in multiple diffraction
signals each linear in the charge density given by n-dimensional correlation
functions of the charge density. A
classical diffraction experiment, in contrast, only reveals even order correlation
functions \cite{ben14}. Since the phase of the charge density in momentum-space
corresponds to translation in real-space, correlation functions such
as $\left\langle \sigma\left(\mathbf{q}_{1},t_1\right)\sigma\left(\mathbf{q}_{2},t_2\right)\right\rangle $ carry interesting structural-dynamical information that is inaccessible
with classical light.

\begin{figure*}[t]
	\centering \includegraphics[width=0.9\textwidth]{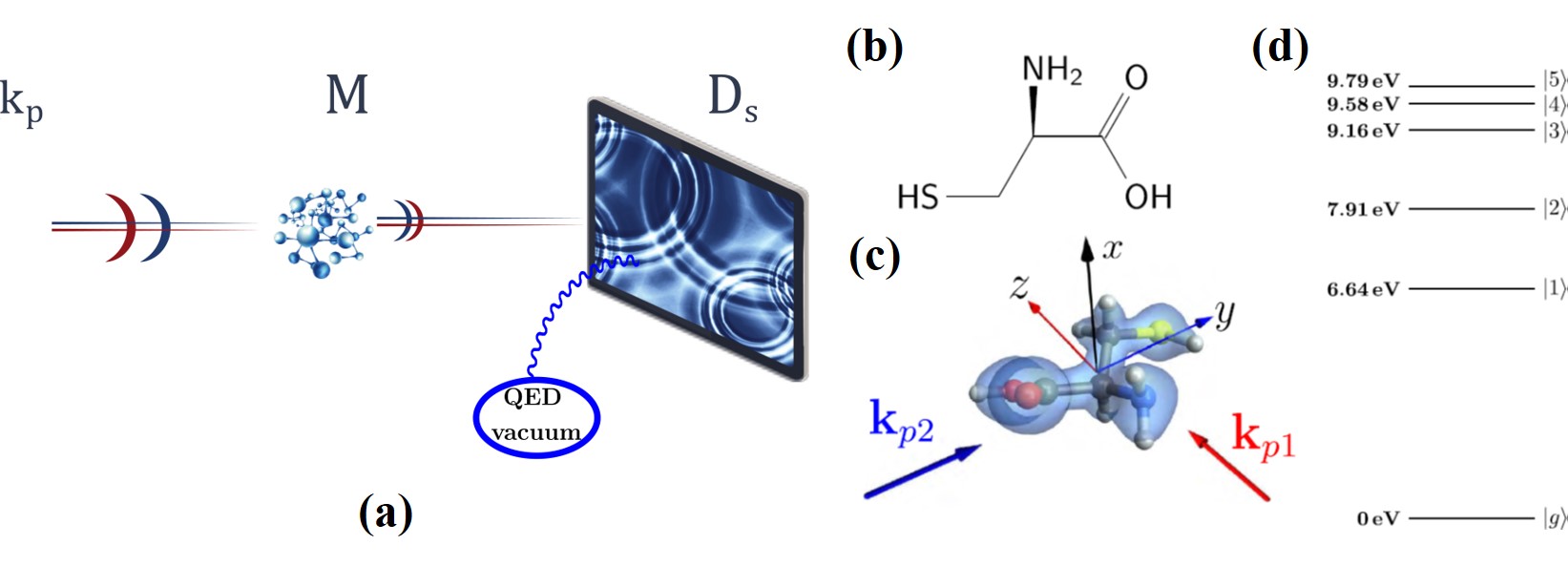}
	\caption{(a) The LQD setup: single photon with momentum $\mathbf{k}_{p}$ diffracted
		off a single molecule and the LQD is detected on a screen (preparation
		pulse is not depicted). The blue circle represents the quantum vacuum fluctuations of QED that interacts once with the detector in the LQD scheme. (b) The chemical structure of cysteine. (c) The orientated cysteine
		and the ground state charge density $\sigma_{gg}$. The pulse configuration:
		$\mathbf{k}_{p1}||\hat{z}$ and $\mathbf{k}_{p2}||\hat{y}$. (d) Energy
		levels of the ground ($g$) and valence excited ($e=1,2,\cdots,5$)
		states.}\label{fig:scheme}
\end{figure*}

\section{The LQD signal} 
Off-resonant diffraction is described by the minimal  coupling matter/field interaction Hamiltonian \cite{tanaka2001time, chernyak2015non}, $\mathcal{H}_{I}=\int d\mathbf{r}~\sigma\left(\mathbf{r},t\right)\mathbf{A}^{2}\left(\mathbf{r},t\right)$
where $\sigma$ is the charge-density operator, while $\mathbf{A}$
is the vector potential.
We first assume that the incoming light pulse is described by a multi-mode coherent
state $\vert\psi_{p}\left(0\right)\rangle=\prod_{\mathbf{p},\lambda}|\alpha_{\mathbf{p},\lambda}\rangle$.
Here $\alpha_{\mathbf{p},\lambda}$ 
represents the amplitude of coherent state of a mode with momentum $\mathbf{p}$
and polarization $\lambda$. The diffraction pattern is obtained from the time-integrated spatially-gated intensity at point $\mathbf{r}$ of the detector. Assuming no temporal gate $\tilde{F}_{t}^{I}(\bar{t},\omega)=2\pi\delta(\omega)$
and performing rotational averaging $\langle\hat{\mathbf{r}}_{m}\hat{\mathbf{r}}_{n}\rangle=\delta_{m,n}/3$,
the first order expansion of the signal in Eq.\,(\ref{eq:S100}) assumes the form
\begin{align}
S_{m}^{(1)}\left[\mathbf{q}_{\left\{ \mathbf{k}\right\} }\left(\mathbf{r}\right)\right]\propto & \mathfrak{Re}\sum_{\mathbf{k},\mathbf{k}_{p}}\omega_{\mathbf{k}}\mathcal{E}_{m}^{*}(\mathbf{k})\mathcal{A}_{m}(\mathbf{k}_{p})\label{eq:Smi1}\\
 & \times\langle\sigma\left[\mathbf{q}_{\left\{ \mathbf{k}\right\} }\left(\mathbf{r}\right),\omega_{\mathbf{q}}\right]\rangle e^{-i\mathbf{\mathbf{q}_{\left\{ \mathbf{k}\right\} }\left(\mathbf{r}\right)}\cdot\mathbf{r}},
\end{align}
 where $m$ is a cartesian component of the field, $\mathbf{q}_{\left\{ \mathbf{k}\right\} }\left(\mathbf{r}\right)=\mathbf{k}_{p}-k\hat{\mathbf{r}}$
and $\omega_{\mathbf{q}}=\omega_{\mathbf{k}_{p}}-\omega_{\mathbf{k}}$
are the diffraction wavector at a corresponding frequency, $\hat{\mathbf{r}}$
is a unit vector in the detection direction; the field and the vector
potential amplitudes $\mathcal{E}_{m}(\mathbf{k})$ and $\mathcal{A}_{n}(\mathbf{k}_{p})$
are given by expectation values of the corresponding operators (see
Eqs.\,(\ref{eq:Em}) and (\ref{eq:An})). The signal
Eq.\,(\ref{eq:Smi1}), which depends on the momentum $\mathbf{k}$, is
governed by the initial state configuration, polarization and other
degrees of freedom. The spatial resolution is controlled by the
the diffraction wavector $\mathbf{q}_{\left\{ \mathbf{k}\right\} }\left(\mathbf{r}\right)$;  $\omega_{\mathbf{q}}$ can be a useful tool for monitoring
transient states of the charge density. A similar result is obtained for a single-photon Fock state $|\psi_{1F}(0)\rangle=\sum_{\mathbf{p},\lambda}\Phi_{\mathbf{p},\lambda}|1_{\mathbf{p},\lambda}\rangle$
(see Appendix \ref{sec:int}), where $\Phi_{\mathbf{p},\lambda}$
represents the Fock state amplitude.

\textit{Time-resolved LQD.} In this setup, an actinic pulse  initially prepares the molecule in a superposition of electronic states and the LQD performed after a delay $T$ probes the excited state dynamics. The superposition of electronic states is described
by density matrix elements $\rho_{ab}^{(0)}$ with the phase $e^{i\phi_{ab}}$,
where $a$ and $b$ are molecular electronic eigenstates. The impulsive diffraction
off this state after time delay $T$ is governed
by the transition charge density element $\sigma_{ab}=\langle a|\hat{\sigma}|b\rangle$. The sum-over-states
expression of Eq. (\ref{eq:Smi1}) for a coherent or single photon
state the reads
\begin{align}
S_{m}^{(1)}&\left[\mathbf{q}\left(\mathbf{r}\right),T\right]\propto \mathfrak{Re}\sum_{\mathbf{k},\mathbf{k}_{p}}\sum_{a,b}\omega_{\mathbf{k}}\mathcal{E}_{m}^{*}(\mathbf{k})\mathcal{A}_{m}(\mathbf{k}_{p})\nonumber \\
 & \times\text{Tr}\left\{ \sigma_{ab}\left[\mathbf{q}\left(\mathbf{r}\right),\omega_{\mathbf{q}}\right]\rho_{ab}^{(0)}\right\} e^{-i\mathbf{q}\left(\mathbf{r}\right)\cdot\mathbf{r}+i\phi_{ab}}.
\end{align}

Thus, the LQD signal may reveal
the single molecule coherence and its phase as well as the transient
charge density and its phase.

We now turn to a different state of the incoming field: an $N$-photon
Fock state described by the wavefunction $|\psi_{NF}(0)\rangle=\sum_{\mathbf{p},\lambda}\Phi_{\mathbf{p},\lambda}^{(N)}|N_{\mathbf{p},\lambda}\rangle$
where $\Phi_{\mathbf{p},\lambda}^{(N)}$ is the $N$-photon amplitude of the $\mathbf{p}$ mode with polarization
$\lambda$. Assuming no temporal gating we obtain from Eq.\,(\ref{eq:S100})
upon oreintational averaging,
\begin{align}
S_{m}^{(1)}\left[\mathbf{q}\left(\mathbf{r}\right)\right]\propto & \mathfrak{Re}\sum_{\mathbf{k}_{p},\lambda}\mathcal{E}_{m\lambda}^{*}(\mathbf{k}_{p})\mathcal{A}_{m\lambda}(\mathbf{k}_{p})\nonumber \\
& \times\langle\sigma\left[\mathbf{q}\left(\mathbf{r}\right),0\right]\rangle e^{-i\mathbf{q}\left(\mathbf{r}\right)\cdot\mathbf{r}},\label{eq:Slin22}
\end{align}
where the abbreviated wavector $\mathbf{q}\left(\mathbf{r}\right)\equiv\mathbf{q}_{\left\{ \mathbf{k}\right\} }\left(\mathbf{r}\right)=\mathbf{k}_{p}-k_{p}\hat{\mathbf{r}}$
and $\mathcal{A}_{m\lambda}(\mathbf{k}_{p})$ are defined in Eqs.
(\ref{eq:Eml}) and (\ref{eq:Anl}), respectively. Note that,
unlike the coherent or the single photon initial states, the $N$-photon
Fock state signal depends solely on the pump momentum and carries no
temporal information, since the frequency argument in the charge density
is zero. This can be explained as follows: The $N$-photon Fock state has a fixed number of photons in each mode. Thus, annihilation and consequent creation of the photon must occur in the same mode to conserve the photon number. In contrast, annihilation of the photon in the single-photon Fock state yields the vacuum state. Thus, the diffracted photon created from the vacuum may have a different momentum. The coherent source has a well-defined average photon number, rather than a fixed photon number, which allows diffraction into a mode other than the pump.
Similarly the time-resolved equivalent of Eq.\,(\ref{eq:Slin22}) yields
\begin{align}
&S_{m}^{(1)}\left[\mathbf{q}\left(\mathbf{r}\right),T\right]\propto \mathfrak{Re}\sum_{\mathbf{k}_{p},\lambda}\sum_{a,b}\mathcal{E}_{m\lambda}^{*}(\mathbf{k}_{p})\mathcal{A}_{m\lambda}(\mathbf{k}_{p})\nonumber \\
& \times\text{Tr}\left\{ \sigma_{ab}\left[\mathbf{q}\left(\mathbf{r}\right),0\right],0)\rho_{ab}^{(0)}(T)\right\} e^{-i\mathbf{q}\left(\mathbf{r}\right)\cdot\mathbf{r}+i\phi_{ab}}.
\end{align}
Heterodyne detection with a classical local oscillator field measures
the interference of a non-interacting local oscillator with interacting
beam (see Eq. (\ref{eq:Smv1})). The role of the local
oscillator is then played by vacuum field fluctuations, which couple to
modes scattered off the matter.

\begin{figure}[t]
	\centering \includegraphics[width=0.45\textwidth]{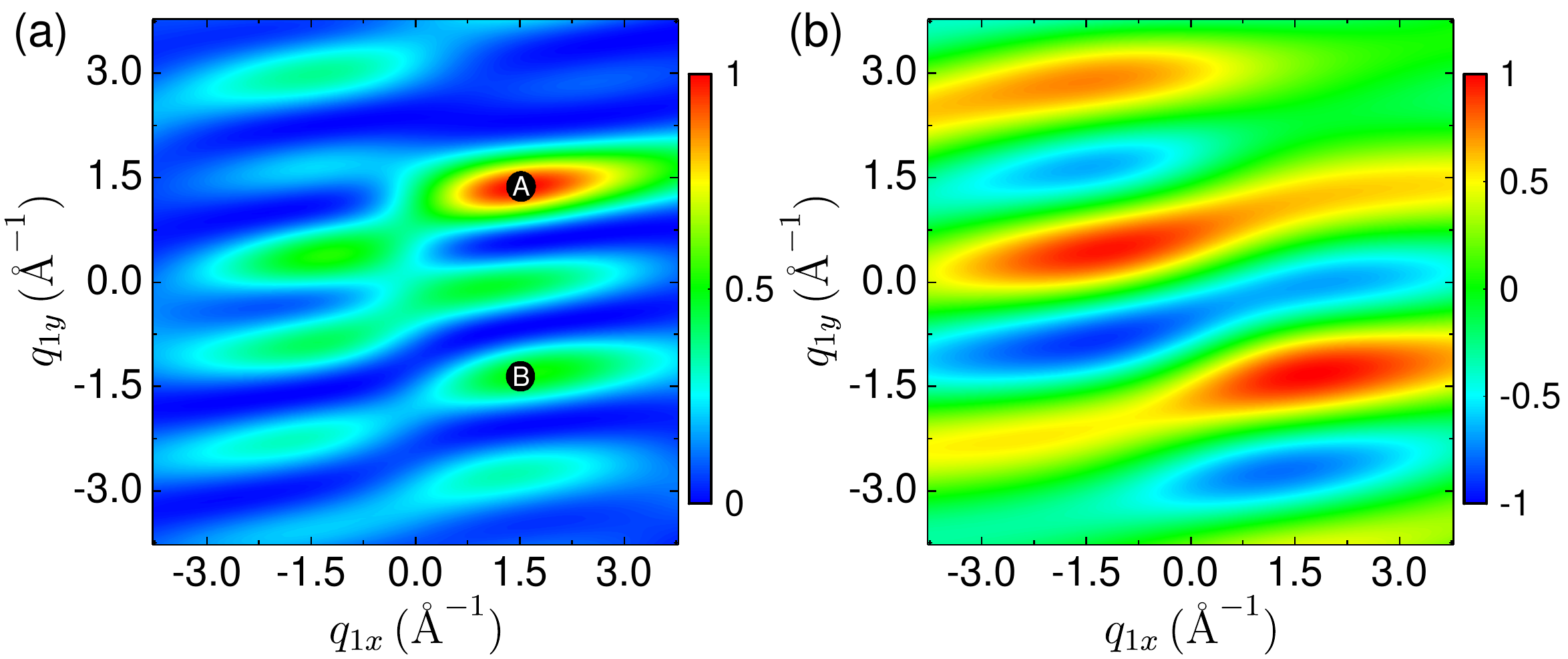}
	\caption{ (a) Classical homodyne and (b) first order linear quantum diffraction
		$\mathbf{q}_{1}$ scattering pattern in the $q_{1z}=1.89\, \text{\AA}^{-1}$
		plane. The first pulse $\mathbf{k}_{p1}$ propagates along $z$. Points
		A and B were used in the calculated diffraction signals.}
	\label{q1}
\end{figure}

\section{Multidimensional quantum diffraction} 
Spontaneous fluctuations of any physical quantity are described by
its multi-point correlation functions. In the case of the charge density
these are $\langle\sigma\left(\mathbf{q}_{1},T_{1}\right)\sigma\left(\mathbf{q}_{2},T_{2}\right)...\rangle$.
We now show how these can be measured by a series of quantum diffraction
processes. We consider a single molecule undergoing
a sequence of \cite{mukamel2015special} n quantum diffraction events. The pulses can have arbitrary
spectral and temporal profiles, provided they are temporally well-separated
and tuned far from any material resonance. 
An $n$-th order coincidence counting of LQD photons
at positions $(\mathbf{r}_{1},\mathbf{r}_{2},...,\mathbf{r}_{n})$
is generated by multiple incoming single photon pulses with momenta
$(\mathbf{k}_{p1},\mathbf{k}_{p2},...,\mathbf{k}_{pn})$ and delays $(T_{1},T_{2},...,T_{n})$.
\begin{align}
S_{q}^{(n)} & \left(\mathbf{q}_{1},T_{1};...;\mathbf{q}_{n},T_{n}\right)\propto\nonumber \\
 & \mathcal{E}_{1}....\mathcal{E}_{n}\times\langle\langle\mathcal{T}\bar{\sigma}\left(\mathbf{q}_{1},T_{1}\right)...\bar{\sigma}\left(\mathbf{q}_{n},T_{n}\right)\rangle,\label{eq:Snq}
\end{align}
where $\bar{\sigma}=\sigma+\sigma^{\dagger}$, and the first momentum transfer is $\mathbf{q}_{1}=\mathbf{k}_{p1}-\mathbf{k}_{1}$, followed by $\mathbf{q}_{2}=\mathbf{k}_{p2}-\mathbf{k}_{2}$, etc, with $\mathbf{k}_{n}$
being the wavevector of the scattered photon.

\begin{figure*}[t]
	\centering \includegraphics[width=0.8\textwidth]{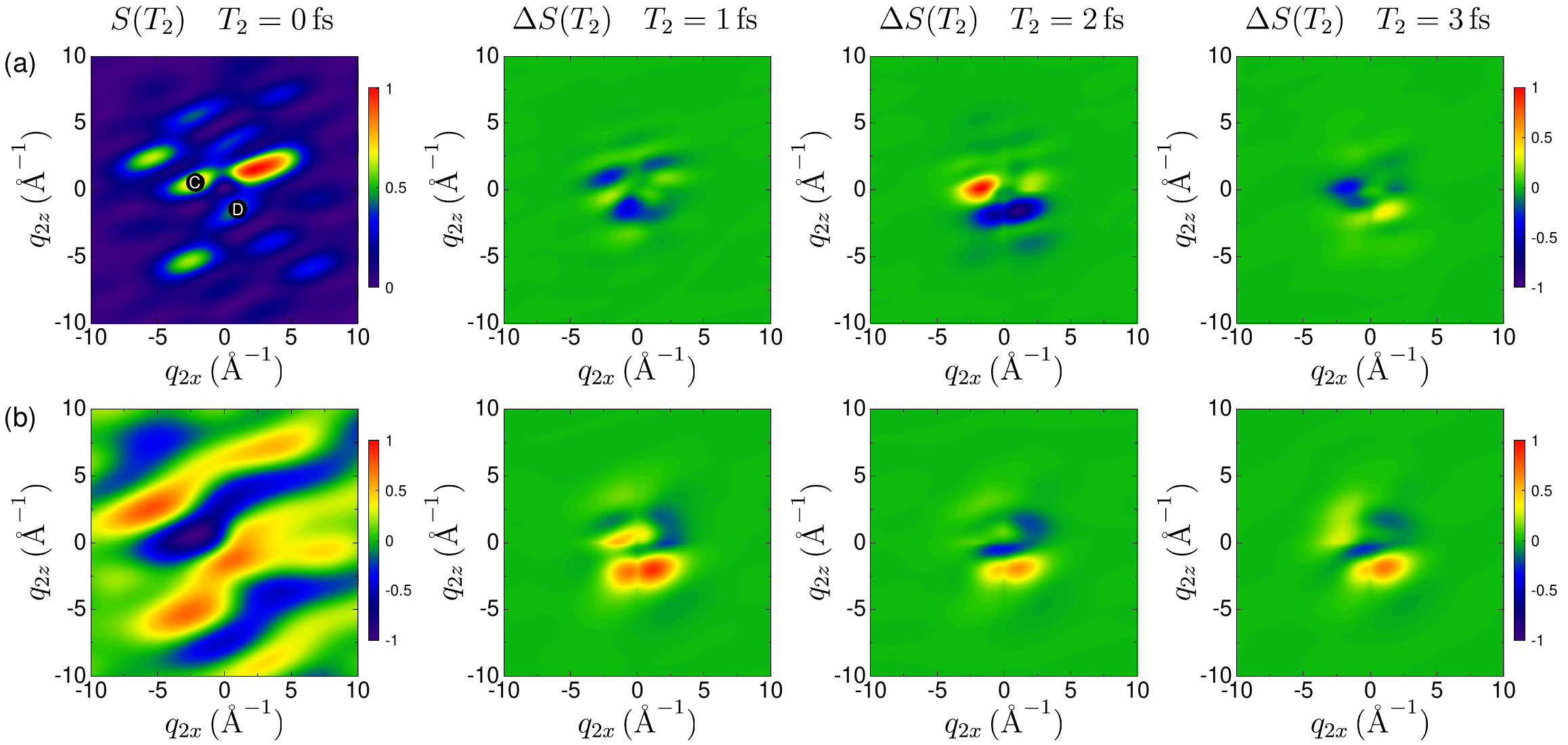}
	\caption{ The $\mathbf{q}_{2}$ diffraction patterns at the given $\mathbf{q}_{1}$
		point A for (a) classical homodyne $S_{c}^{(4)}(\mathbf{q}_{1},T_{1}=0;\mathbf{q}_{2},T_{2})$
		and (b) second order linear quantum diffraction $S_{q}^{(2)}(\mathbf{q}_{1},T_{1}=0;\mathbf{q}_{2},T_{2})$
		signals at four time delays $T_{2}$ in the $q_{2y}=1.89\, \text{\AA}^{-1}$
		plane. The second pulse $\mathbf{k}_{p2}$ propagates along $y$.
		To highlight the changes, in columns 2, 3, and 4, we plot the signal
		difference $\Delta S(T_{2})=S(T_{2})-S(T_{2}=0)$. }
	\label{q2}
\end{figure*}

We now examine the two lowest-order signals. In the simplest (2D) experiment,
the molecule is subjected to two off-resonant pulses, with wavevectors
$\mathbf{k}_{p1}$ and $\mathbf{k}_{p2}$. A scattered single-photon
amplitude from pulse $1$ with frequency $\omega_{1}$ is contracted
with the incoming photon amplitude, and the resulting photon is collected
in the direction $\mathbf{k}_{1}$ at time $T_{1}$. The molecule
is in a superposition state during the interpulse delay, after
which the second pulse is scattered, and the photon amplitude is contracted
with the incoming photon amplitude such that the resulting photon
with frequency $\omega_{2}$ is collected in the $\mathbf{k}_{2}$
direction at time $T_{2}$.



In the impulsive limit, the 2D signal can be written as (see Fig. \ref{S2-loop} and discussion therein)
\begin{align}
S^{(2)}(\mathbf{q}_{1},T_{1};\mathbf{q}_{2},T_{2}) & \propto\langle\bar{\sigma}(\mathbf{q}_{2},T_{2})\bar{\sigma}(\mathbf{q}_{1},T_{1})\rangle.\label{eq:S2qt}
\end{align}
Higher-order signals can be calculated similarly.



We have simulated the 2D diffraction signals Eq.\,(\ref{eq:S2qt})
from a single oriented cysteine molecule (see Fig.\,\ref{fig:scheme}(b)
and (c)). Quantum chemistry calculations were performed by using the
MOLPRO code \cite{werner2010molpro}. The optimized geometry was obtained
at the Hartree-Fock/cc-pVDZ \cite{dunning1989gaussian} level of theory.
The lowest six   valence electronic energy levels were calculated at the CASSCF(6/6)/cc-pVDZ
level of theory \cite{werner1980quadratically,werner1985second,knowles1985efficient} are depicted in Fig.\,\ref{fig:scheme}(d).
The transition density matrix was evaluated using
\begin{align}
\sigma_{ij}(\mathbf{r})=\sum_{mn}T_{mn}^{(ij)}\phi_{m}(\mathbf{r})\phi_{n}(\mathbf{r}).
\end{align}
Here, the indices $i$, $j$ run over the valence eigenstates. $T_{mn}^{(ij)}$
is the transition density-matrix element between states $i$
and $j$ for the $m$ and $n$ atomic orbitals.

We consider diffraction signals from the ground state ($g$). In the
impulsive limit, the classical homodyne signal for single pulse scattering
is given by the sum-over-states expression Eq.\,(\ref{S2c}) which determines the (transition) charge density $\sigma_{ag}(\mathbf{q}_{1})$
between two electronic states $a$ and $g$ in momentum $\mathbf{q}_{1}\equiv\mathbf{k}_{1}-\mathbf{k}_{p1}$
space. The homodyne detected signal Eq.\,(\ref{S2c}) misses the phase of
$\sigma_{ag}(\mathbf{q}_{1})$.

The  linear (1D) quantum diffraction signal solely gives the
ground state charge density:
\begin{align}
S_{q}^{(1)}(\mathbf{q}_{1},T_{1})\propto & \left\langle \bar{\sigma}(\mathbf{q}_{1},T_{1})\right\rangle =2\mathfrak{Re}[ \sigma_{gg}(\mathbf{q}_{1},T_1)] .\label{S1}
\end{align}
Both signals Eqs.\,(\ref{S2c}) and (\ref{S1}) are independent on the time delay $T_{1}$. Time-dependent signals can be obtained by first preparing the molecule
in a superposition state \cite{ben14,PhysRevLett.120.243902}. The first pulse $\mathbf{k}_{p1}$ propagates along $z$, while the
$\mathbf{q}_{1}$ diffraction signals $S_{c}^{(2)}(\mathbf{q}_{1})$
and $S_{q}^{(1)}(\mathbf{q}_{1})$ are detected in the $(q_{1x},q_{1y})$
plane; see Fig.\,\ref{q1}. The scattering shows rich pattern
in $\mathbf{q}_{1}$ space. The homodyne detected signal (Fig.\,\ref{q1}(a))
is positive, and several peaks can be observed in the $\mathbf{q}_{1}$
domain. The linear quantum diffraction signal may be negative. 
The classical $S_{c}^{(2)}$ signal (Eq.\,(\ref{S2c})) is expressed
as the modulus square form of (transition) charge densities in momentum
space. The quantum $S_{q}^{(1)}$ signal (Eq.\,(\ref{S1})) in contrast depends
on both the amplitude and phase of charge densities, making it possible
to extract the real space ground state charge density $\sigma_{gg}(\mathbf{r})$.

\begin{figure*}[t]
	\centering \includegraphics[width=0.8\textwidth]{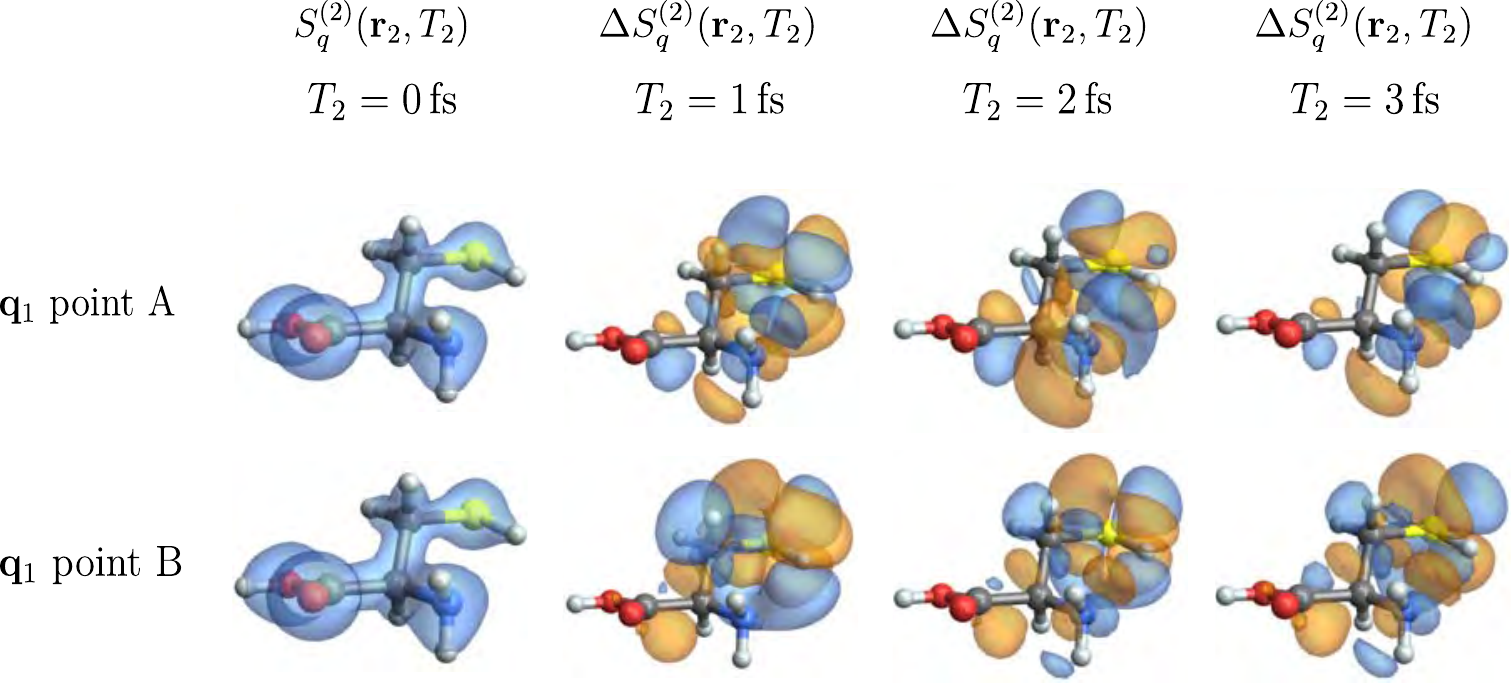}
	\caption{The time-dependent charge density obtained using Eq.\,(\ref{Sr2})
		for $\mathbf{q}_{1}$ points ${\rm A}=(1.51\, \text{\AA}^{-1},1.36\, \text{\AA}^{-1})$
		and ${\rm B}=(1.51\, \text{\AA}^{-1},-1.36\, \text{\AA}^{-1})$ marked
		in Fig.\,\ref{q1}. We show the full signal at $T_{2}=0\,{\rm fs}$,
		and the signal differences $\Delta S_{q}^{(2)}(\mathbf{r}_{2},T_{2})=S_{q}^{(2)}(\mathbf{q}_{1},T_{1}=0;\mathbf{r}_{2},T_{2})-S_{q}^{(2)}(\mathbf{q}_{1},T_{1}=0;\mathbf{r}_{2},T_{2}=0)$
		for other plots. }
	\label{sigma_T}
\end{figure*}

To study the charge density dynamical fluctuations, we resort to the $S_{q}^{(2)}$
signal.
\noindent To investigate the two-photon coincidence scattering pattern
in $\mathbf{q}_{2}$ space, we select the $\mathbf{q}_{1}$ point ${\rm A}=(1.51\, \text{\AA}^{-1},1.36\, \text{\AA}^{-1})$
in Fig.\,\ref{q1}(a). The classical homodyne signal for $\mathbf{q}_{2}$
scattering is given by Eq.\,(\ref{S4c_1}). 
The second order LQD signal is
\begin{align}
 & S_{q}^{(2)}(\mathbf{q}_{1},T_{1}=0;\mathbf{q}_{2},T_{2})\propto\left\langle \bar{\sigma}(\mathbf{q}_{2},T_{2})\bar{\sigma}(\mathbf{q}_{1},0)\right\rangle \nonumber \\
= & \sum_{a}[\sigma_{ga}(\mathbf{q}_{2})+\sigma_{ag}^{\ast}(\mathbf{q}_{2})][\sigma_{ag}(\mathbf{q}_{1})+\sigma_{ga}^{\ast}(\mathbf{q}_{1})]e^{-i\omega_{ag}T_{2}}.\label{S2}
\end{align}
Because the molecule is initially in the ground state, the signal  depends only on the second time delay $T_{2}$.

Figure\,\ref{q2} depicts the $\mathbf{q}_{2}$ scattering pattern
in the $q_{2y}=1.89\, \text{\AA}^{-1}$ plane, where the second pulse
$\mathbf{k}_{p2}$ propagates along $y$. The first column shows the
signals at $T_{2}=0$. Again, we see that the classical $S_{c}^{(4)}$
signal (Fig.\,\ref{q2}(a)) is always positive, while the quantum
signal $S_{q}^{(2)}$ (Fig.\,\ref{q2}(b)) may be negative. Since
signal is  dominated by  the time-independent pathways, \textit{i.e.}, $c=d$ in Eq.\,(\ref{S4c_1})
and $a=g$ in Eq.\,(\ref{S2}), the diffraction
signals at different time delays look very similar. To better visulize
the changes, we plot the signal difference $S(T_{2})-S(T_{2}=0)$
in columns 2, 3, and 4, where the time-independent background has been subtracted.
Rich temporal patterns in $\mathbf{q}_{2}$ originates
from interferences between the various scattering pathways.
By Fourier transform of the time-domain signal into the frequency
($\Omega$) domain, we can identify the electronic coherences that
contribute to the dynamics of the signal. If Fig. \ref{signal_freq} of the SI we display such spectra at the points C and D of Fig. \ref{q2}.

The classical $S_{c}^{(2)}$ signal represents the electron density
fluctuations in momentum space, and may be used to image the real-space
charge-density correlation functions. The quantum phase-dependent
$S_{q}^{(2)}$ signal, in contrast, can retrieve the time-dependent
transition charge densities in real space. Fourier transformation of the
second order LQD signal Eq.\,(\ref{S2}) into real space $\mathbf{r}_{2}$
at a given $\mathbf{q}_{1}$ point in Fig.\,\ref{q1} gives
\begin{align}
 & S_{q}^{(2)}(\mathbf{q}_{1},T_{1}=0;\mathbf{r}_{2},T_{2})  \nonumber \\
\propto & \int d\mathbf{q_{2}}\,e^{i\mathbf{q}_{2}\cdot\mathbf{r}_{2}}  S_{q}^{(2)}(\mathbf{q}_{1},T_{1}=0;\mathbf{q}_{2},T_{2})
\nonumber \\
= & 2\sum_{a}[\sigma_{ag}(\mathbf{q}_{1})+\sigma_{ga}^{\ast}(\mathbf{q}_{1})]e^{-i\omega_{ag}T_{2}}\sigma_{ga}(\mathbf{r}_{2}).\label{Sr2}
\end{align}
Figure\,\ref{sigma_T} depicts the real-space signal at the two $\mathbf{q}_{1}$
points A and B marked in Fig.\,\ref{q1} for different time delays $T_{2}$.
At $T_{2}=0\,{\rm fs}$, the signal looks similar to the ground state
charge density (see Fig.\,\ref{fig:scheme} (c)), because the $\sigma_{gg}(\mathbf{r}_{2})$
term dominates Eq.\,(\ref{sigma_T}). As in Fig. 3 the other plots at $T_{2}\ne0$
have this signal subtracted, and thus image the dynamics of transition
charge densities in real space. For points A and B in Fig.$\left(\ref{q1}\right)$,
the real-space signals show a very different time dependence, because
different momenta $\mathbf{q}_{1}$ are transferred to the electrons
by the first pulse $\mathbf{k}_{p1}$. The spatial Fourier-transformed
real-space signal Eq.\,(\ref{Sr2}) is a combination of various (transition)
charge densities $\sigma_{ga}(\mathbf{r})$, and can provide information
about quantum coherence between the ground and excited states.

\section{Discussion}

To compare the LQD signals Eq.\,(\ref{eq:Snq}) with classical diffraction \cite{kow17} we first note that the former vanishes for a classical
field and requires a quantum field. Furthermore, using
a light source in which the quantum nature of radiation is prominent, the
signal reveals both the amplitude and phase of the charge density.
The diffraction can originate from a group of molecules initially
in their ground states with a small fraction in the excited state.
The relevant material quantity in Eq. (\ref{eq:S2c1}) is then $\langle\sigma_{gg}\rangle_{\alpha}\langle\sigma_{ag}\rangle_{\beta}$
where $\alpha$ and $\beta$ represent two molecules,  $\sigma_{ag}$ is
a transition charge density (coherence). Classical homodyne diffraction
is quadratic in the charge density and originates from pairs of molecules
(see Appendix \ref{sec:homc}). The single-molecule contribution,
in contrast, originates solely from excited state population since
the trace of the diffracted field operators in the expectation value
with respect to vacuum state of the field vanishes if the molecule
is in a coherent superposition. Classical diffraction carries no information about single molecule coherence.

The multidimensional extension of diffraction imaging with classical light
to $n$ diffraction events scales to $n$-th order in the light \textit{intensity}
and $2n$-th order in the charge density (see Appendix \ref{sec:homc}). The corresponding quantum light signal presented here, in contrast,
scales to $n$-th order in the field \textit{amplitude}. Thus, at a given intensity quantum
light allows to observe higher order correlations, thanks to the more
favorable intensity scaling. Classical homodyne diffraction dominated by even  orders in the charge density is governed by the static (localized)
charge density while the new information carried by the phase in the
odd contributions provides a novel way of measuring transient charge
density, density-density correlations and dynamical events in molecules using quantum diffraction.
Generally, the $n$-th order signals have both amplitude square contributions
and lower order phase dependent contributions (such as the ones explored
in Fig.\,(\ref{fig:scheme})). For intense quantum sources with
many photons the contribution quadratic in the charge density dominates and the phase dependent
terms merely provide a minor correction to the strong background.
It is therefore critical to use low photon fluxes in order to isolate
the phase-dependent contributions. An alternative way to single out these
terms is by employing multiple single photon interferences generated
by introducing beam splitters in e.g. Mach-Zehnder interferometers (MZI). The phase of the classical local oscillator field allows to separate real and imaginary part of the material response function and extract the phase in heterodyne measurement. Similar results can be obtained for quantum field by combining the MZI with the phase plates. The
multidimensional analogue will be an interesting topic for a future study.

\section{Acknowledgements}

K.E.D. is supported by the Zijiang Endowed Young Scholar
Fund and Overseas
Expertise Introduction Project for Discipline Innovation (111 Project,
B12024). S.M. acknowledges the National Science Foundation (grant
CHE-1361516) and the support of the Chemical Sciences, Geosciences,
and Biosciences division, Office of Basic Energy Sciences, Office
of Science, U.S. Department of Energy through award No. DE-FG02-04ER15571 and DE-SC0019484.
S.A. was supported by the DOE grant. We wish to thank Noa Asban
for the graphical illustrations.

\appendix
\begin{widetext}

\section{Derivation of the LQD signal}

\begin{figure*}[t]
	\centering \includegraphics[width=0.8\textwidth]{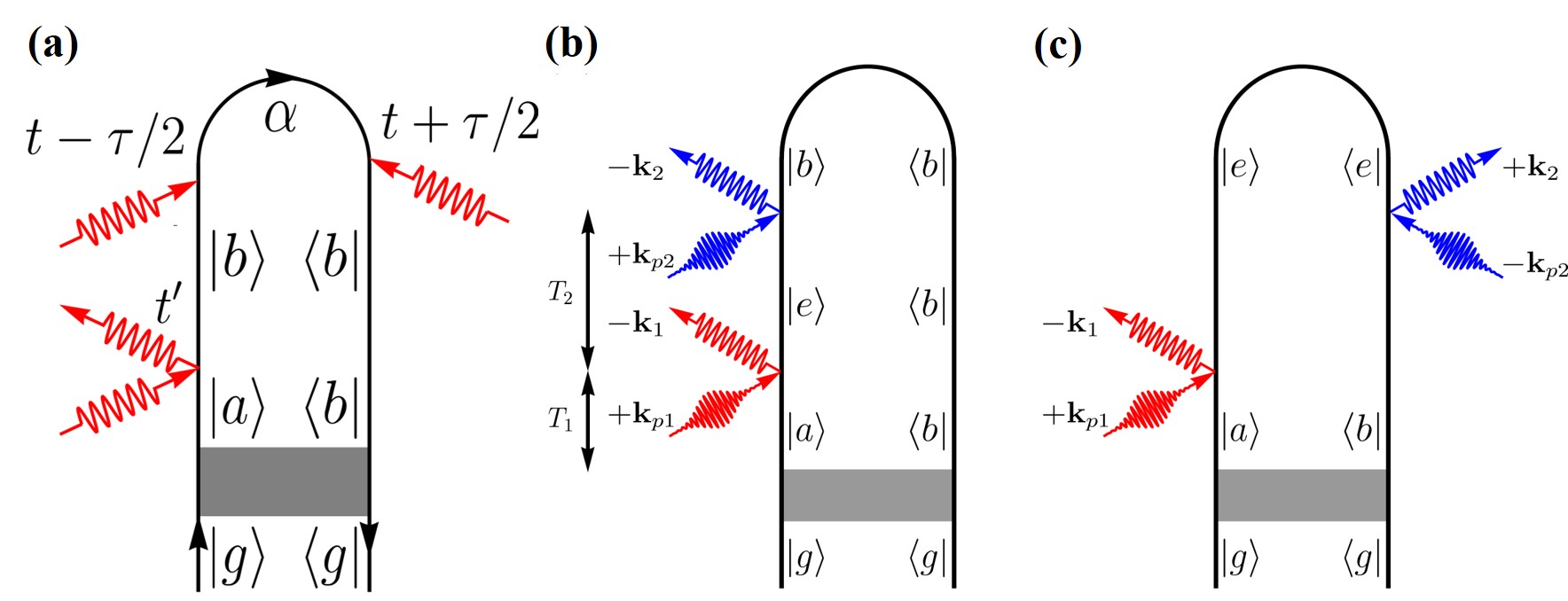}
	\caption{	(a) The loop diagram for the LQD process. An actinic
		pulse (shaded area) prepares the molecule in a superposition of electronic
		states $\rho_{ab}$. After the LQD the state of the system is $\rho_{bb}$.\
		The top two arrows represent the detection.(b) and (c) The loop diagrams representing the LQD signal in Eq. (\ref{eq:S2qt})
		resulting from two successive scattering measurements. For diagram rules see
		Ref. \cite{mar08}}
	\label{S2-loop}
\end{figure*}

\label{sec:int}

The diffraction pattern is obtained from the time-integrated spatially-gated intensity at point $\mathbf{r}$ of the detector.
\begin{align}
& S_{m}\left[\mathbf{q}\left(\mathbf{r}\right)\right]=\int dtF_{t}^{I}(\bar{t},t)\nonumber \\
& \times\left\langle \mathcal{T}\mathbf{E}_{m}^{\left(-\right)}\left(\mathbf{r},t\right)\mathbf{E}_{m}^{\left(+\right)}\left(\mathbf{r},t\right)e^{-\frac{i}{\hbar}\int_{-\infty}^{t}d\tau\mathcal{H}_{I-}(\tau)}\right\rangle ,\label{eq:S100}
\end{align}
where $m$ is cartesian component of the field, $\mathbf{q}\left(\mathbf{r}\right)$ is the diffraction wavevector
corresponing to detection at point \textbf{$\mathbf{r}$}, $F_{t}^{I}(\bar{t},t)$
is a temporal gate, and $\mathcal{T}$ is the time ordering superoperator.
$\mathcal{H}_{I-}$ is the interaction superoperator, defined by its action on an
ordinary operator $X$ according to $\mathcal{H}_{I-}X\equiv\mathcal{H}_{I}X-X\mathcal{H}_{I}$
\cite{Har08}. 
By expanding the exponent to first order in $\mathcal{H}_{I-}$, and
separating the incoming (pump) from the detected modes of the electric
field we obtain the LQD signal (see Eq.\,(\ref{eq:Slin0})).
For brevity we assume a temporal gating $F_{t}^{I}(\bar{t},t)$ that
acts on the intensity, rather than the field (which was $F_{t}(\bar{t},t)$).
The first order expansion of Eq. (\ref{eq:S100}) in field-matter
interaction yields:
\begin{align}
S_{m}^{(1)}(\mathbf{r}) & =\frac{2}{\hbar}\mathfrak{Im}\int dtF_{t}^{I}(\bar{t},t)\int_{-\infty}^{t}dt'\int d\mathbf{r}'\langle\sigma(\mathbf{r}',t')\rangle_{\mu}\notag\\
&\times\sum_{n}\langle\psi_{p}(0)|\mathbf{E}_{m}^{(-)}(\mathbf{r},t)\mathbf{A}_{n}^{(+)}(\mathbf{r}',t')|\psi_{p}(0)\rangle\langle0|\mathbf{E}_{m}^{(+)}(\mathbf{r},t)\mathbf{A}_{n}^{(-)}(\mathbf{r}',t')|0\rangle,\label{eq:Slin0}
\end{align}
where $\mathfrak{Im}$ denotes imaginary part, $n$ represents the cartesian
coordinates of the vector potential coming from $\mathbf{A}^{2}$
interaction term and $\langle...\rangle_{\mu}\equiv\text{Tr}[...\rho_{\mu}^{(0)}]$
is taken with respect to the initial state of the molecule $\rho_{\mu}^{(0)}$,
and $\psi_{p}(0)$ is the state of the pump photon source.

While the second correlation function over the vacuum state representing
detection modes is the same as in the field amplitude signal,
the first correlation function over the pump photon state is more
peculiar. Assuming the coherent state, the field correlation function
reads
\begin{align}
& \langle\psi_{p}(0)|\mathbf{E}_{m}^{(-)}(\mathbf{r},t)\mathbf{A}_{n}^{(+)}(\mathbf{r}',t')|\psi_{p}(0)\rangle=\mathcal{E}_{m}^{*}(\mathbf{r},t)\mathcal{A}_{n}(\mathbf{r}',t'),\label{eq:2ptpump}
\end{align}
where $\mathcal{E}_{m}(\mathbf{r},t)=\sum_{\mathbf{k},\mu}\sqrt{\frac{2\pi\hbar\omega_{\mathbf{k}}}{V_{\mathbf{k}}}}\epsilon_{m}^{(\mu)}(\mathbf{k})\alpha_{\mathbf{k},\mu}e^{i(\mathbf{k}\cdot\mathbf{r}-\omega_{\mathbf{k}}t)}$
and $\mathcal{A}_{n}(\mathbf{r},t)=-ic\sum_{\mathbf{k},\mu}\sqrt{\frac{2\pi\hbar}{\omega_{\mathbf{k}}V_{\mathbf{k}}}}\epsilon_{m}^{(\mu)}(\mathbf{k})\alpha_{\mathbf{k},\mu}e^{i(\mathbf{k}\cdot\mathbf{r}-\omega_{\mathbf{k}}t)}$
with $\alpha_{\mathbf{k},\mu}=\langle\alpha|\hat{a}_{\mathbf{k},\mu}|\alpha\rangle$.
Note that a similar expression can be achieved for a single photon
Fock state $|\psi_{1F}(0)\rangle=\sum_{\mathbf{p},\lambda}\Phi_{\mathbf{p},\lambda}|1_{\mathbf{p},\lambda}\rangle$.
In this case the corresponding field amplitudes are given by $\mathcal{E}_{m}(\mathbf{r},t)=\sum_{\mathbf{k},\mu}\sqrt{\frac{2\pi\hbar\omega_{\mathbf{k}}}{V_{\mathbf{k}}}}\epsilon_{m}^{(\mu)}(\mathbf{k})\Phi_{\mathbf{k},\mu}e^{i(\mathbf{k}\cdot\mathbf{r}-\omega_{\mathbf{k}}t)}$
and $\mathcal{A}_{n}(\mathbf{r},t)=-ic\sum_{\mathbf{k},\mu}\sqrt{\frac{2\pi\hbar}{\omega_{\mathbf{k}}V_{\mathbf{k}}}}\epsilon_{m}^{(\mu)}(\mathbf{k})\Phi_{\mathbf{k},\mu}e^{i(\mathbf{k}\cdot\mathbf{r}-\omega_{\mathbf{k}}t)}$
with $\Phi_{\mathbf{k},\mu}=\langle0|\hat{a}_{\mathbf{k},\mu}|\psi_{1F}(0)\rangle$.
Following the method outlined previously and using the following identity:

\begin{align}
& \frac{2\pi\hbar}{V_{\mathbf{k}}}\sum_{\mathbf{k}}\epsilon_{m}^{(\mu)}(\mathbf{k})\epsilon_{n}^{(\mu)}(\mathbf{k})e^{i\mathbf{k}\cdot\mathbf{R}}\int_{-\infty}^{t}dt'e^{-i\omega_{\mathbf{k}}(t-t')-i\Omega t'}=\nonumber \\
& \frac{\hbar}{4\pi}\left[\frac{\Omega^{2}}{c^{2}}(\delta_{m,n}-\hat{\mathbf{r}}_{m}\hat{\mathbf{r}}_{n})+\left(\frac{i\Omega}{cR}-\frac{1}{R^{2}}\right)(\delta_{m,n}-3\hat{\mathbf{r}}_{m}\hat{\mathbf{r}}_{n})\right]\frac{e^{-i\Omega(t-R/c)}}{\Omega R}
\end{align}
we obtain for the signal
\begin{align}
& S_{m}^{(1)}(\mathbf{r})\propto\mathfrak{Re}\int\frac{d\omega'}{2\pi}\sum_{\mathbf{k},\mathbf{k}_{p}}\tilde{F}_{t}^{I}(\bar{t},\omega'-\omega_{\mathbf{k}_{p}}+\omega_{\mathbf{k}})\langle\sigma(\mathbf{q}_{r}(\omega'),\omega')\rangle\sum_{n}\mathcal{E}_{m}^{*}(\mathbf{k})\mathcal{A}_{n}(\mathbf{k}_{p})\nonumber \\
& \times\left[\frac{\omega_{\mathbf{k}_{p}}-\omega'}{c}(\delta_{m,n}-\hat{\mathbf{r}}_{m}\hat{\mathbf{r}}_{n})+\frac{i}{r}(\delta_{m,n}-3\hat{\mathbf{r}}_{m}\hat{\mathbf{r}}_{n})\right]e^{i(\omega_{\mathbf{k}_{p}}-\omega')r/c-i\mathbf{k}\cdot\mathbf{r}}
\end{align}
where $\tilde{F}_{t}^{I}(\bar{t},\omega)=\int dte^{i\omega t}F_{t}^{I}(\bar{t},t)$
is a Fourier transform of the gating function,
\begin{align}
\mathcal{E}_{m}^{*}(\mathbf{k})=\sum_{\mu}\sqrt{\frac{2\pi\hbar\omega_{\mathbf{k}}}{V_{\mathbf{k}}}}\epsilon_{m}^{(\mu)}(\mathbf{k})\chi_{\mathbf{k},\mu},\label{eq:Em}
\end{align}
\begin{align}
\mathcal{A}_{n}(\mathbf{k}_{p})=-ic\sum_{\lambda}\sqrt{\frac{2\pi\hbar}{\omega_{\mathbf{k}_{p}}V_{\mathbf{k}_{p}}}}\epsilon_{n}^{(\lambda)}(\mathbf{k}_{p})\chi_{\mathbf{k}_{p},\lambda},\label{eq:An}
\end{align}
where $\chi=\alpha$ for coherent state and $\chi=\Phi$ for single
photon Fock state. Now assuming no temporal gate and taking rotating
averaging we obtain Eq. (\ref{eq:Smi1}).

For the $N$-photon Fock state the field correlation function Eq.\,(\ref{eq:2ptpump})
will contain only same momentum and polarization components of the
two fields:
\begin{align}
& \langle\psi_{p}(0)|\mathbf{E}_{m}^{\dagger}(\mathbf{r},t)\mathbf{A}_{pn}(\mathbf{r}',t')|\psi_{p}(0)\rangle=\sum_{\mathbf{k}_{p},\lambda}\mathcal{E}_{m\lambda}^{*}(\mathbf{k}_{p})\mathcal{A}_{n\lambda}(\mathbf{k}_{p})e^{-i[\mathbf{k}_{p}(\mathbf{r}-\mathbf{r}')-\omega_{\mathbf{k}_{p}}(t-t')]},
\end{align}
where
\begin{align}
\mathcal{E}_{m\lambda}^{*}(\mathbf{k}_{p})=\sqrt{\frac{2\pi\hbar\omega_{\mathbf{k}_{p}}N_{\mathbf{k}_{p},\lambda}}{V_{\mathbf{k}_{p}}}}\Phi_{\mathbf{k}_{p},\lambda}^{(N)*}\epsilon_{m}^{(\lambda)}(\mathbf{k}_{p}),\label{eq:Eml}
\end{align}
and
\begin{align}
\mathcal{A}_{n\lambda}(\mathbf{k}_{p})=-ic\sqrt{\frac{2\pi\hbar N_{\mathbf{k}_{p},\lambda}}{\omega_{\mathbf{k}_{p}}V_{\mathbf{k}_{p}}}}\Phi_{\mathbf{k}_{p},\lambda}^{(N)}\epsilon_{n}^{(\lambda)}(\mathbf{k}_{p}).\label{eq:Anl}
\end{align}
Here we use  $\langle\psi_{NF}(0)|\hat{a}_{\mathbf{k},\nu}^{\dagger}\hat{a}_{\mathbf{k}_{p},\lambda}|\psi_{NF}(0)\rangle=N_{\mathbf{k}_{p},\lambda}|\Phi_{\mathbf{k}_{p},\lambda}^{(N)}|^{2}\delta_{\mathbf{k},\mathbf{k}_{p}}\delta_{\nu,\lambda}$.
Following the outlined approach the LQD signal yields
\begin{align}
& S_{m}^{(1)}(\mathbf{r})\propto\mathfrak{Re}\int\frac{d\omega'}{2\pi}\sum_{\mathbf{k}_{p},\lambda}\tilde{F}_{t}^{I}(\bar{t},\omega')\langle\sigma(\mathbf{q}_{r}(\omega'),\omega')\rangle\sum_{n}\mathcal{E}_{m\lambda}^{*}(\mathbf{k}_{p})\mathcal{A}_{n\lambda}(\mathbf{k}_{p})\nonumber \\
& \times\left[\frac{\omega_{\mathbf{k}_{p}}-\omega'}{c}(\delta_{m,n}-\hat{\mathbf{r}}_{m}\hat{\mathbf{r}}_{n})+\frac{i}{r}(\delta_{m,n}-3\hat{\mathbf{r}}_{m}\hat{\mathbf{r}}_{n})\right]e^{i(\omega_{\mathbf{k}_{p}}-\omega')r/c-i\mathbf{k}_{p}\cdot\mathbf{r}},
\end{align}
where $\mathbf{q}\left(\mathbf{r}\right)=\mathbf{k}_{p}+\frac{\omega'-\omega_{p}}{c}\hat{\mathbf r}$.
Assuming no temporal gate and taking rotating averaging we simplify
the signal to Eq.\,(\ref{eq:Slin22}).

The 2D extension of the LQD signal  resulting from the two successive
scattering measurements is described by the two diagrams of Fig.$\left(\ref{S2-loop}\right)$
(and their complex conjugates) stemming from the separation of $\sigma$
and $\sigma^{\dagger}$ that correspond either to both scattering
events occur with the ket, or with the ket and the bra (see Fig.\,\ref{S2-loop}(b)).
The complex conjugate diagrams are not shown. The signal can be read off the diagram and is given by Eq. (\ref{eq:S2qt}).

\section{Diffraction of classical light}\label{sec:homc}

\begin{figure*}[t]
	\centering \includegraphics[width=0.8\textwidth]{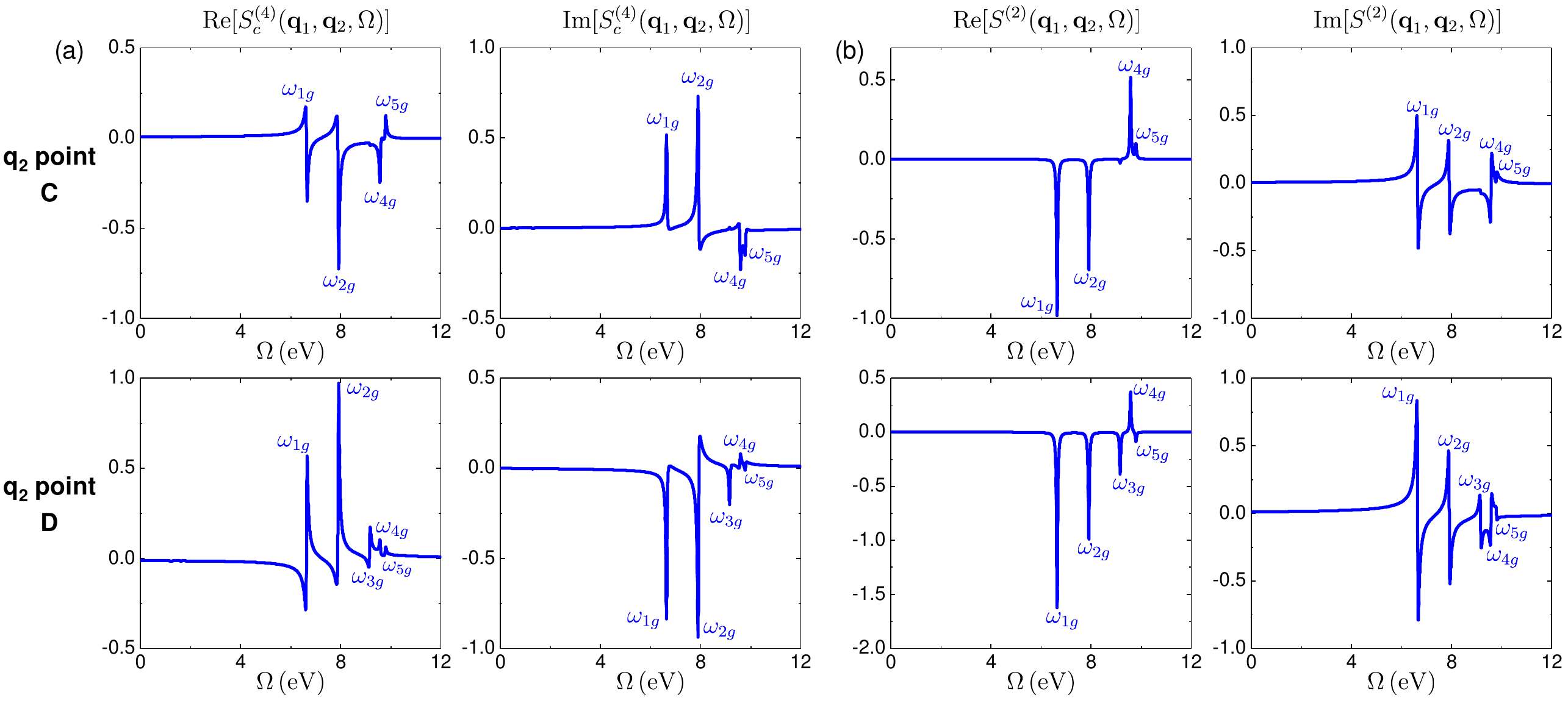}
	\caption{ The real and imaginary parts of signals for (a) classical homodyne
		$S_{c}^{(4)}(\mathbf{q}_{1},\mathbf{q}_{2},\Omega)$ (columns 1 and
		2) and (b) LQD $S_{q}^{(2)}(\mathbf{q}_{1},\mathbf{q}_{2},\Omega)$
		(columns 3 and 4) in frequency domain. The first and second rows show
		signals at the $\mathbf{q}_{2}$ points C and D marked in Fig. \ref{q2}(a),
		respectively. }
	\label{signal_freq}
\end{figure*}

\subsection{Heterodyne detection}

\label{sec:hetc} A classical heterodyne diffraction is measured by
mixing classical diffracted field with another classical local oscillator
field. The signal is given by
\begin{align}
S_{c}(\mathbf{r})=\frac{2}{\hbar}\mathfrak{Im}\int dt\mathcal{A}_{d}^{*}(\mathbf{r},t)\int dt_{1}d\mathbf{r}_{1}\mathcal{A}_{p}(\mathbf{r}_{1},t_{1})\langle\sigma(\mathbf{r}_{1},t_{1})\rangle_{\mu}
\label{eq:Smv1}
\end{align}
This signal is also linear in the charge density. Following the similar
derivation presented in Appendix A we obtain for the signal in the
CW limit
\begin{align}
& S_{c}^{(1)}(\mathbf{r})\propto\mathfrak{Re}[\mathcal{A}_{d}^{*}(\omega_{0})\mathcal{A}_{p}(\omega_{0})\langle\sigma\left(\mathbf{q}_{\mathbf r}(0),0\right)\rangle_{\mu}e^{-i\mathbf{k}_{p'}\cdot\mathbf{r}}].\label{eq:Sclin1}
\end{align}

\subsection{Homodyne detection}
Unlike quantum case and heterodyne classical detection, classical homodyne signal is
linear in the field intensity and quadratic in the charge density.
\begin{align}
S_{c}^{(2)} & (\mathbf{r})=\frac{2}{\hbar^{2}}\mathfrak{Re}\int dtd\mathbf{r}_{1}d\mathbf{r}_{1}dt_{1}dt_{2}\langle\sigma\left(\mathbf{r}_{2},t_{2}\right)\sigma\left(\mathbf{r}_{1},t_{1}\right)\rangle_{\mu}\nonumber \\
& \times\langle \mathbf{A}^{2}\left(\mathbf{r}_{2},t_{2}\right)\mathbf{E}^{\left(-\right)}\left(\mathbf{r},t\right)\mathbf{E}^{\left(+\right)}\left(\mathbf{r},t\right)\mathbf{A}^{2}\left(\mathbf{r}_{1},t_{1}\right)\rangle_{\phi}.
\end{align}

Following the same steps discussed above the homodyne signal for CW
pump is given by
\begin{align}
& S_{c}^{(2)}(\mathbf{r})\propto\langle|\sigma\left(\mathbf{q}_{\mathbf r}(0),0\right)|^{2}\rangle_{\mu}.\label{eq:S2c1}
\end{align}
Assuming no preparation and time-delayed diffraction denoted by $T_{1}$
and expanding the signal Eq.\,(\ref{eq:S2c1}) in sum-over states yields
\begin{align}
S_{c}^{(2)}(T_{1},\mathbf{q}_{1})\propto & \left\langle \sigma^{\dagger}(T_{1},\mathbf{q}_{1})\sigma(T_{1},\mathbf{q}_{1})\right\rangle =\sum_{a}\left|\sigma_{ag}(\mathbf{q}_{1})\right|^{2}.\label{S2c}
\end{align}

\subsection{Multidimensional classical diffraction}

\label{sec:multc}

The signals studied by Biggs et al. \cite{big14} employ contributions
linear in the field intensities and are analogous to classical signals
which in the impulsive limit read
\begin{align}
& S_{c}^{(2n)}(\mathbf{q}_{1},T_{1};...; \mathbf{q}_{n},T_{n})\propto I_{1}....I_{n}\nonumber \\
& \times\langle\langle\mathcal{T}\sigma^{\dagger}(T_{1},\mathbf{q}_{1})\sigma(T_{1},\mathbf{q}_{1})...\sigma^{\dagger}(T_{n},\mathbf{q}_{n})\sigma(T_{n},\mathbf{q}_{n})\rangle\rangle,\label{eq:Snc}
\end{align}
where $I_{j}=|\mathcal{E}_{j}|^{2}$ are field intensities. In Eq.\,(\ref{eq:Snc})
each diffraction event is quadratic in $\sigma$ and we omitted the
frequency argument in the charge density. 
Expanding the $n=2$ signal in sum-over states yields
\begin{align}
& S_{c}^{(4)}(T_{1}=0,\mathbf{q}_{1},T_{2},\mathbf{q}_{2})\propto\left\langle \sigma^{\dag}(T_{1}=0,\mathbf{q}_{1})\sigma^{\dag}(T_{2},\mathbf{q}_{2})\sigma(T_{2},\mathbf{q}_{2})\sigma(T_{1}=0,\mathbf{q}_{1})\right\rangle \nonumber \\
= & \sum_{ecd}\sigma_{cg}(\mathbf{q}_{1})\sigma_{dg}^{\ast}(\mathbf{q}_{1})\sigma_{ec}(\mathbf{q}_{2})\sigma_{ed}^{\ast}(\mathbf{q}_{2})e^{-i\omega_{cd}T_{2}}.
\label{S4c_1}
\end{align}


\section{The classical homodyne and LQD signals in frequency domain}

\label{sec:classh}

We select two points ${\rm C}=(-2.17\, \text{\AA}^{-1},0.47\, \text{\AA}^{-1})$
and ${\rm D}=(0.95\, \text{\AA}^{-1},-1.42\, \text{\AA}^{-1})$ in Fig.
\ref{q2}(a), and depict the frequency-domain signals in Fig.\,\ref{signal_freq}.
The peaks at $\Omega=0$, which correspond to the time-independent
pathways and overwhelm other peaks away from the origin ($\Omega=0$), 
have been excluded from Fig.\,\ref{signal_freq}. We conclude that
the time dependence of the signal is determined by the energy differences
between the ground and the excited states.

\end{widetext}


%

\end{document}